\documentclass[sigconf,nonacm]{acmart}
\usepackage{tabularx}
\makeatletter
\@ACM@balancefalse
\makeatother
\AtBeginDocument{%
  }

\begin{document}

\title{Westlake Scholar: AI-Enhanced Scholarly Discovery over an Institutional Repository}


\author{Junshu Pan}
\authornote{These authors contributed equally to this work.}
\author{Luodan Zhang}
\authornotemark[1]
\author{Yifeng Lu}
\authornotemark[1]
\affiliation{%
  \department{School of Engineering}
  \institution{Westlake University}
  \city{Hangzhou}
  \country{China}
}

\author{Mengfan Zhao}
\author{Ming Luo}
\affiliation{%
  \department{Office of Information Technology (Library)}
  \institution{Westlake University}
  \city{Hangzhou}
  \country{China}
}

\author{Zijie Yang}
\correspondingauthor
\email{yangzijie@airalogy.com}
\affiliation{%
  \institution{Airalogy}
  \city{Hangzhou}
  \country{China}
}
\affiliation{%
  \department{School of Engineering}
  \institution{Westlake University}
  \city{Hangzhou}
  \country{China}
}

\author{Yue Zhang}
\correspondingauthor
\email{zhangyue@westlake.edu.cn}
\affiliation{%
  \department{School of Engineering}
  \institution{Westlake University}
  \city{Hangzhou}
  \country{China}
}

\author{Rui Shang}
\correspondingauthor
\email{shangrui@westlake.edu.cn}
\affiliation{%
  \department{Office of Information Technology (Library)}
  \institution{Westlake University}
  \city{Hangzhou}
  \country{China}
}

\begin{abstract}
Institutional repositories (IRs) provide mature infrastructure for preserving and disseminating research outputs, but conventional record- and document-centric interfaces provide limited support for connecting deposited papers to related research and people. We present Westlake Scholar, an open-source, institution-grounded platform that adds four complementary artificial intelligence (AI) services to repository infrastructure: contextual paper reading, research-direction-guided paper discovery, publication-grounded expert discovery, and AI-generated research chronologies for scholars. The services draw on a shared institutional knowledge layer connecting approved publication records, paper content, and scholar--publication relationships. This allows the same paper to support contextual reading, cross-paper discovery, expert matching, and longitudinal views of scholarly work.

Westlake Scholar provides an open and governable implementation of an institution-controlled AI layer that connects repository content, scholarly discovery, and researcher relationships while preserving provenance, human review, and institutional governance. A deployment at Westlake University, in operation since April 2026, demonstrates that the integrated system can operate in a live institutional setting.

\end{abstract}

\keywords{institutional repositories, library and information science, AI-assisted scholarly discovery, research information management, institutional governance}



\maketitle

\begin{figure}[t]
  \centering
  \anon[{\includegraphics[width=\columnwidth]{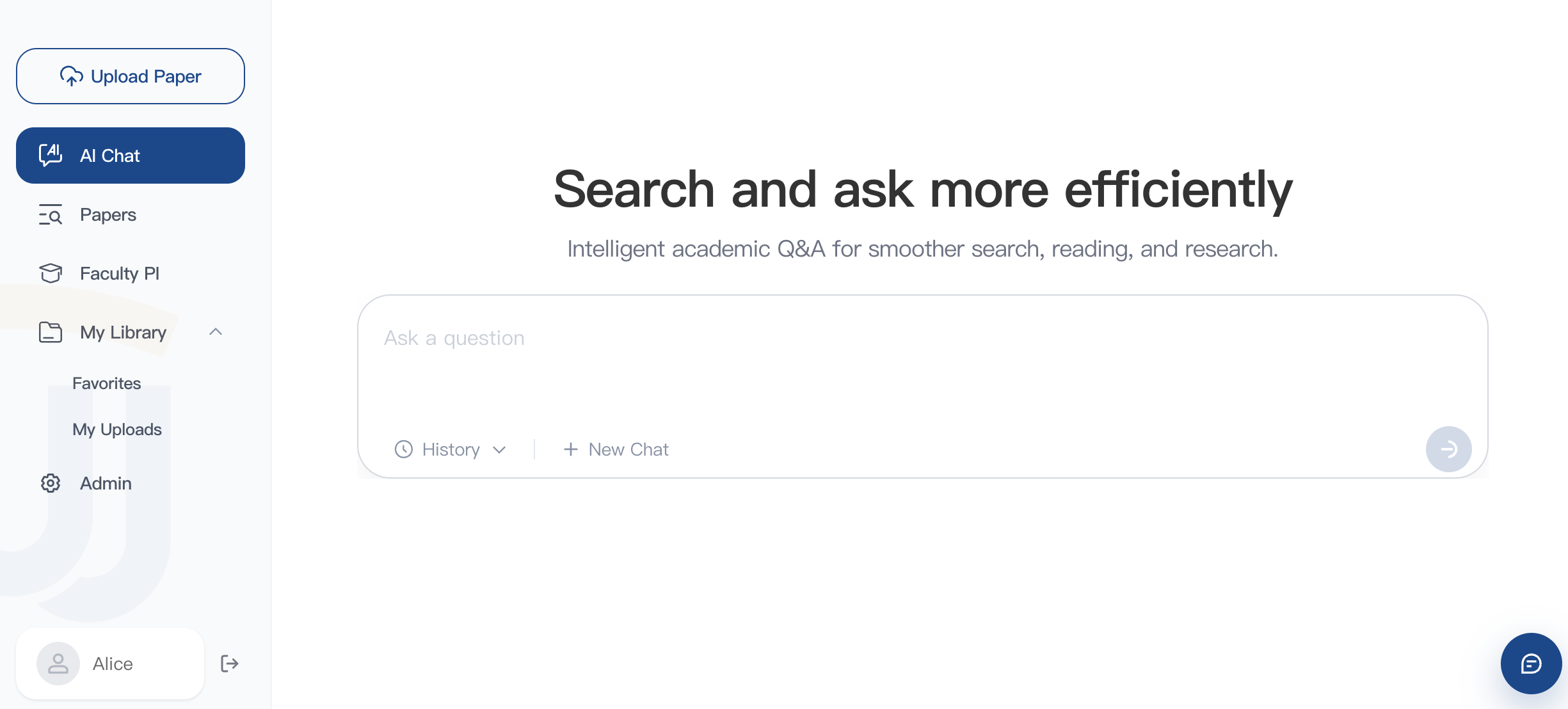}}]{\includegraphics[width=\columnwidth]{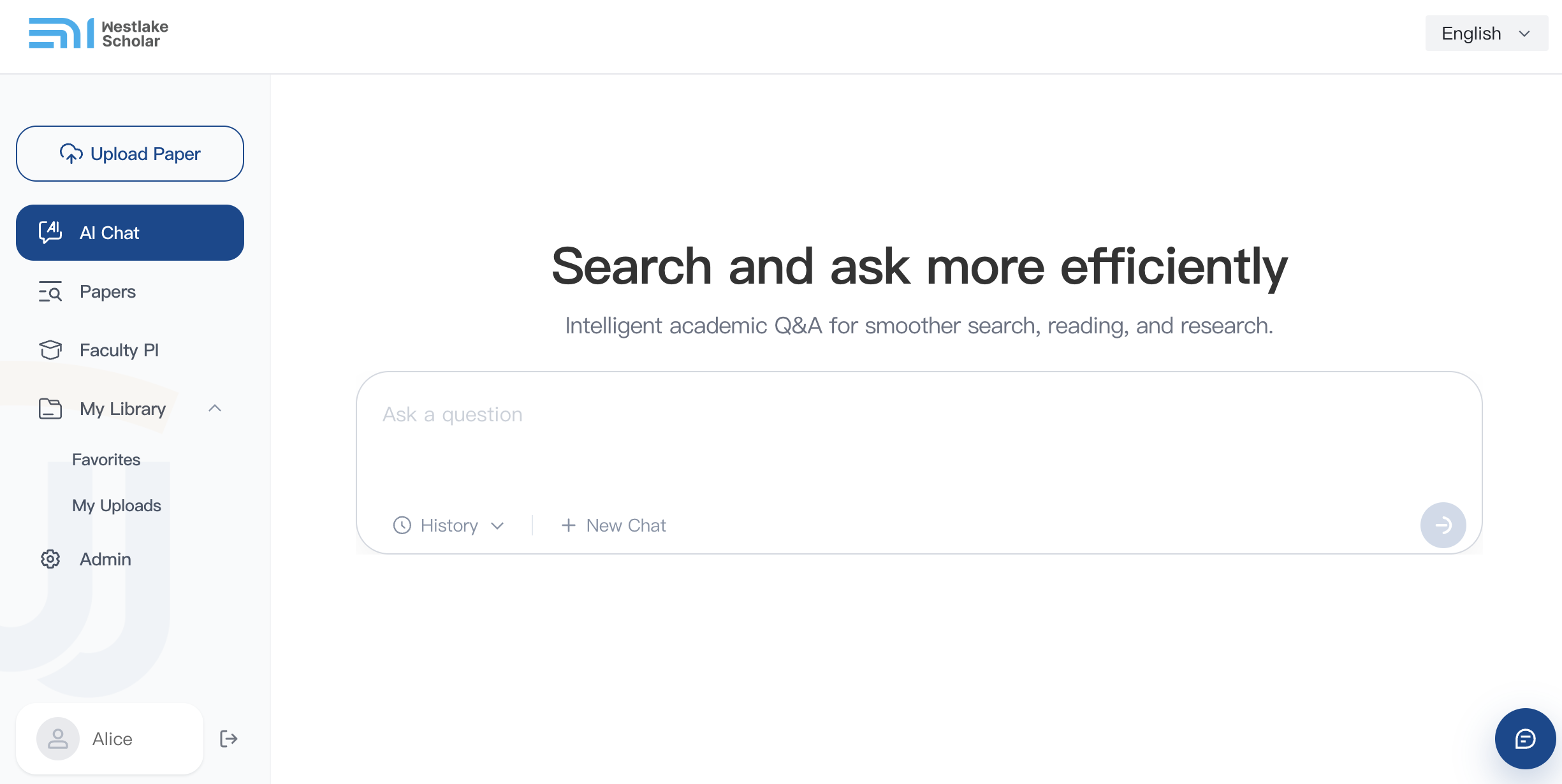}}
  \caption{\anon[The system]{Westlake Scholar} integrates repository browsing, conversational assistance, and scholar discovery.}
  \label{fig:system-overview}
  \Description{The \anon[anonymous system]{Westlake Scholar} interface provides navigation to paper upload, AI chat, papers, institutional expert discovery, and a personal library, with a central conversational search area.}
\end{figure}

\section{Introduction}
Institutional repositories (IRs) are foundational infrastructure for collecting, preserving, and disseminating research outputs~\cite{Lynch2003}. Their dominant access model, however, remains centered on metadata records, document lists, and individual files. Making research outputs available through an IR does not by itself ensure their discoverability or effective use~\cite{MacGregor2019, Arlitsch2022, Davis2024}. At institutional scale, this post-deposit gap appears in four connected needs: interpreting an unfamiliar paper, tracing relationships across a growing collection, locating relevant expertise across organizational boundaries, and understanding how a scholar's interests evolve.

Retrieval-augmented language models can connect questions to repository evidence~\cite{Asai2026}. Scholarly graphs make relations among works, authors, institutions, and concepts machine actionable~\cite{Jaradeh2019, Priem2022}. The relevant design question is therefore not whether an IR can add a generic chatbot, but how an institution-controlled data layer can support several related scholarly tasks while preserving source context, organizational scope, and content governance.

We introduce Westlake Scholar (Figure~\ref{fig:system-overview}), which combines repository functions for depositing, cataloging, searching, browsing, and accessing scholarly outputs with four AI-enhanced services over a shared institutional foundation. Its scenario-driven co-development brought together library service expertise, product engineering, and AI research. We make three contributions:

\begin{itemize}
\item \textbf{A shared institution-grounded knowledge foundation.} Approved publication records, paper content, and scholar--publication relationships provide a common foundation for all four services.
\item \textbf{An integrated progression of scholarly discovery.} Building on this foundation, the system connects contextual paper reading, cross-paper discovery, publication-grounded expert matching, and longitudinal interpretation, enabling users to move from a single paper to related work, institutional expertise, and research trajectories.
\item \textbf{Institutional governance across the service lifecycle.} Across these services, institutions control content approval, model processing, and chronology publication, while provenance and human review support inspection and correction.
\end{itemize}

\section{Related Work}

\subsection{Institutional Repositories Beyond Deposit}

Research on IRs has examined participation, service design, and impact~\cite{Arlitsch2021}. Faculty self-archiving is shaped by individual and institutional factors, including researchers' awareness and attitudes~\cite{Kim2011, Creaser2010}. Service-oriented, user-centered, and participatory approaches position repositories as evolving research infrastructure rather than static deposit endpoints~\cite{Armstrong2014, JohnsonFreeman2024, Davis2024, Lankes2008}. Discoverability beyond repository interfaces nevertheless remains a persistent concern, including web visibility and external indexing~\cite{MacGregor2019, Arlitsch2022}.

Research information management (RIM) extends beyond deposit by aggregating and curating information about research across institutional systems~\cite{Bryant2017}. These functions are commonly supported by current research information systems (CRIS) and research information management systems (RIMS). Commercial CRIS/RIMS platforms such as Pure, Symplectic Elements, and Converis connect publications, profiles, projects, reporting, and institutional workflows~\cite{ElsevierPure2026, Symplectic2026, ClarivateConveris2026}. VIVO similarly provides an open semantic infrastructure for representing and discovering researchers, scholarly works, expertise, and organizational relationships~\cite{Borner2012}. These systems integrate research records, profiles, workflows, and scholarly relationships at institutional scale. Westlake Scholar extends this institutional information environment toward interactive post-deposit use: an approved repository corpus and scholar--publication links jointly support contextual paper reading, cross-paper discovery, publication-grounded expert matching, and longitudinal research chronologies.

\subsection{AI-Assisted Scholarly Discovery}

Library-native AI discovery increasingly combines natural-language inquiry with source-linked synthesis. Primo Research Assistant translates natural-language questions into searches over the Central Discovery Index, reranks results, and synthesizes an answer from selected abstracts~\cite{ExLibrisPrimo2026}; implementation studies discuss both its discovery potential and concerns about metadata and recommendation accuracy~\cite{LiWilson2025, Kotula2026}. OpenScholar retrieves from a large open-access literature corpus to produce citation-backed answers~\cite{Asai2026}, while STORM combines retrieval, multi-perspective question asking, and structured synthesis for long-form knowledge curation~\cite{Shao2024}. A library and information science perspective further suggests that AI support should respond to the stages and uncertainties of information seeking~\cite{Kuhlthau1991, Ravuri2025}.

Table~\ref{tab:positioning} identifies Westlake Scholar's system-level contribution. The platform brings retrieval-augmented interaction, semantic discovery, expert recommendations, and research chronologies into an institution-operated service architecture in which approved repository records and scholar links provide a shared grounding layer across all four tasks.

\begin{table*}[t]
  \caption{Comparison of Westlake Scholar with related system families.}
  \label{tab:positioning}
  \small
  \begin{tabularx}{\textwidth}{@{}p{0.16\textwidth}p{0.24\textwidth}p{0.25\textwidth}X@{}}
    \toprule
    \textbf{System family} & \textbf{Data foundation} & \textbf{Primary user-facing focus} & \textbf{Relationship to Westlake Scholar} \\
    \midrule
    Library AI discovery assistants~\cite{ExLibrisPrimo2026, LiWilson2025} & Vendor or library discovery indexes and source descriptions/abstracts & Natural-language literature discovery and source-linked synthesis & Uses broad discovery indexes for conversational search and source-linked synthesis; Westlake Scholar grounds discovery in institution-approved records and connects papers to local scholars and research chronologies. \\
    CRIS/RIMS (Pure, Elements, Converis)~\cite{Bryant2017, ElsevierPure2026, Symplectic2026, ClarivateConveris2026} & Integrated institutional research records, profiles, projects, and workflows & Research administration, reporting, profiles, visibility, and collaboration & Integrates institutional records, profiles, projects, and workflows for research administration; Westlake Scholar centers interactive post-deposit use of repository papers. \\
    VIVO~\cite{Borner2012} & Institutionally maintained linked scholarly data & Scholarly networking, semantic search, expert and relationship discovery & Represents scholarly entities and relationships for semantic discovery; Westlake Scholar connects institutional scholarly relations to conversational paper discovery and evidence-linked research chronologies. \\
    \textbf{Westlake Scholar} & Approved repository records, extracted paper text, semantic indexes, and scholar--publication links & Contextual reading, paper discovery, expert matching, and research chronologies & Reuses one institution-controlled paper layer across all four workflows, with content review and chronology publication controls. \\
    \bottomrule
  \end{tabularx}
\end{table*}

\section{System Design and Implementation}

\subsection{From Institutional Data to Need-Driven AI Services}

Westlake Scholar was co-developed by Westlake University Library, Hangzhou Airalogy Technology Co., Ltd., and the university's AI laboratory. The library defined service scenarios and provided institutional data; Airalogy led product design and engineering; and the AI laboratory contributed AI expertise. The public implementation uses a web application and service API backed by institution-controlled storage and hybrid lexical--vector retrieval. A deployment is scoped to one institution: approved repository records, paper indexes, scholar--publication links, and chronology records remain inside that institutional boundary. Only approved paper claims enter public search and recommendation paths.

\begin{figure*}[t]
  \centering
  \includegraphics[width=\textwidth]{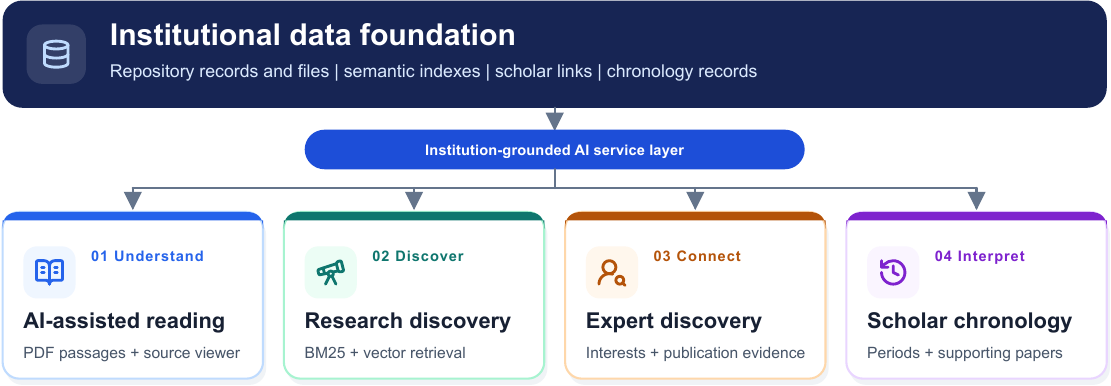}
  \caption{Four task-specific workflows over a shared institutional data foundation.}
  \label{fig:implemented-architecture}
  \Description{A shared institutional data foundation feeds four separate workflows for paper reading, research-direction-guided paper discovery, institutional expert discovery, and research chronologies for scholars.}
\end{figure*}

Figure~\ref{fig:implemented-architecture} summarizes four workflows corresponding to a progression of institutional needs: understanding one paper in context, discovering relationships across the collection, locating relevant expertise, and interpreting a scholar's trajectory. Each interface invokes a predefined task handler---paper-context chat, paper retrieval, scholar recommendation, or chronology generation. The handlers share records and relationships while retaining task-specific prompts and candidate limits, forming a task-routed AI service layer that couples institutional evidence with workflow-specific behavior.

Following the interface-consistency principle commonly summarized as Jakob's Law~\cite{Yablonski2024}, the system retains familiar repository conventions---persistent navigation, search and browse entry points, record views, and an inline document reader---while placing conversational actions alongside them. Users can therefore transfer expectations learned from other information systems rather than relearn the repository interface, consistent with human-centered design guidance to align systems with users' existing practices~\cite{ISO9241}.

\subsection{Retrieval and Model Configuration}

As an institutional repository, Westlake Scholar holds deposited papers alongside their bibliographic metadata. This creates an institution-grounded evidence base in which retrieval can draw on approved paper content as well as titles, abstracts, and authorship records. During indexing, the system extracts text from an approved paper's PDF and combines it with the title and abstract. The resulting text is divided into 1,000-character segments with 100-character overlap. Each segment stores normalized term frequencies for BM25 ranking~\cite{Robertson1995Okapi, Robertson2009}. Full-text content remains within the institutional deployment by default and supports local lexical retrieval; only titles and abstracts are sent for embedding, and paper-context questions use bibliographic metadata. If an institution authorizes its configured model provider to process approved paper content, Westlake Scholar also sends approved text segments to that provider for embedding and supplies retrieved passages to the model for paper-context questions. Lexical ranking prioritizes exact and repeated query terms, while vector retrieval broadens recall to conceptually related language. The paper-search API exposes either retrieval mode. Conversational paper recommendation combines the top results from both modes using reciprocal-rank fusion and returns the best segment from each of three papers. The expert-discovery mode separately uses recent conversational context to retrieve publication--scholar candidates.

Westlake Scholar accesses generative and embedding models through an OpenAI-compatible API. The Westlake University deployment uses \texttt{qwen3.7-plus} for chat and chronology generation, while the public release defaults to \texttt{text-embedding-v4} for embeddings. Institutions can select other compatible chat and embedding models and configure a separate chronology model when needed. Model-provider processing of approved full-text content is disabled by default and requires explicit administrator authorization.

\subsection{AI-Assisted Reading of Deposited Papers}

\begin{figure*}[t]
  \centering
  \includegraphics[width=\textwidth]{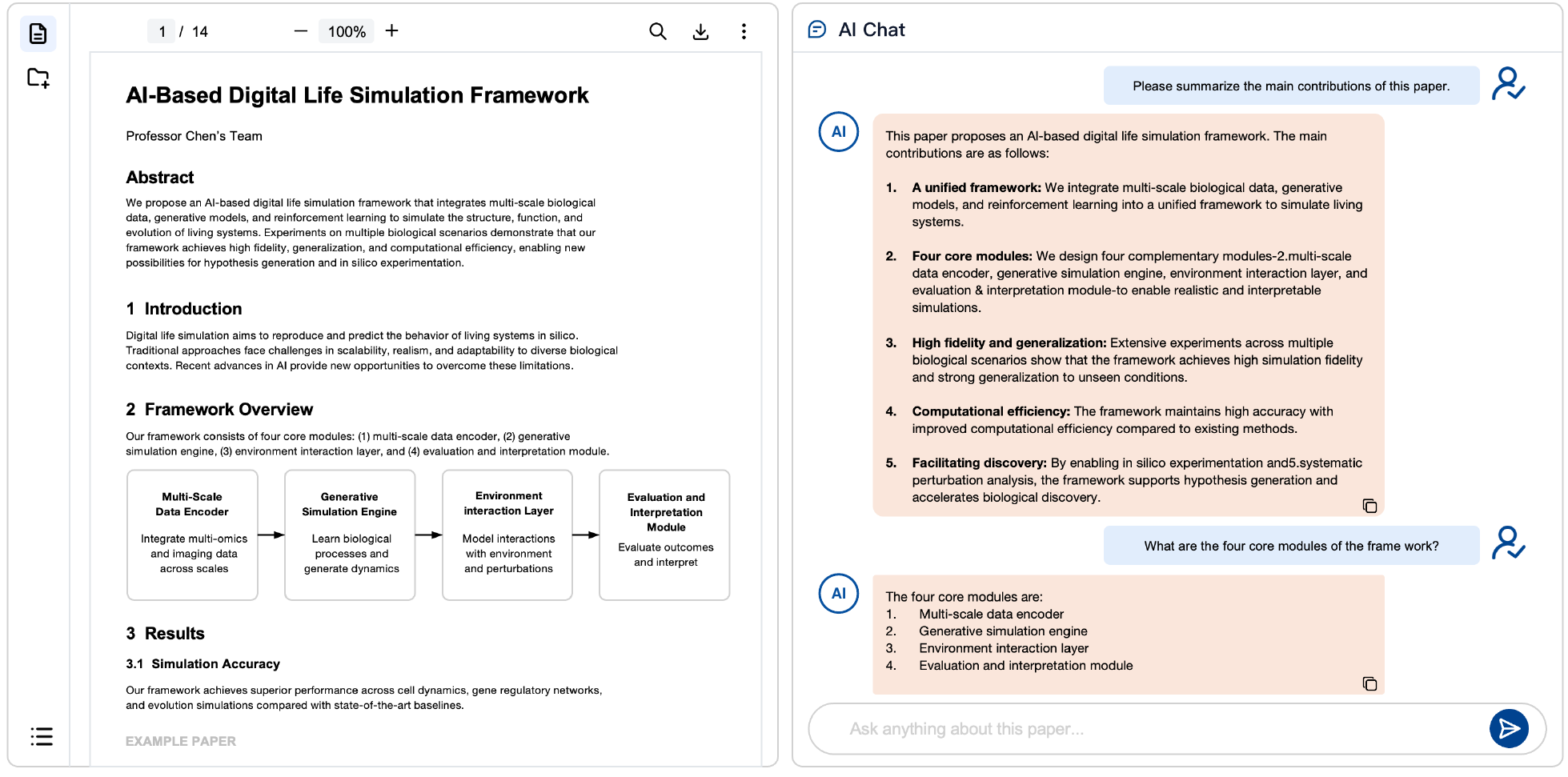}
  \caption{AI-assisted reading keeps a deposited paper and its contextual conversation in one workspace.}
  \label{fig:paper-reading}
  \Description{A deposited paper is displayed beside a conversational assistant that summarizes it and answers a follow-up question about its core modules.}
\end{figure*}

Depositing a paper makes it accessible but does not by itself help readers interpret or connect it to other institutional work. Westlake Scholar places an inline PDF viewer beside a conversational panel. When a reader opens an item, the first turn supplies the assistant with the paper identifier and bibliographic context. Where the deployment administrator has enabled approved-PDF model processing, the server retrieves relevant passages from the paper's indexed PDF text and supplies them as evidence; otherwise the conversation remains grounded in bibliographic metadata. Subsequent turns reuse the saved dialogue. The visible PDF remains the source against which users can check an answer. The assistant may also invoke the institutional paper-recommendation tool when the conversation raises a related research topic (Figure~\ref{fig:paper-reading}).

\subsection{Research-Direction-Guided Paper Discovery}

\begin{figure*}[t]
  \centering
  \includegraphics[width=\textwidth]{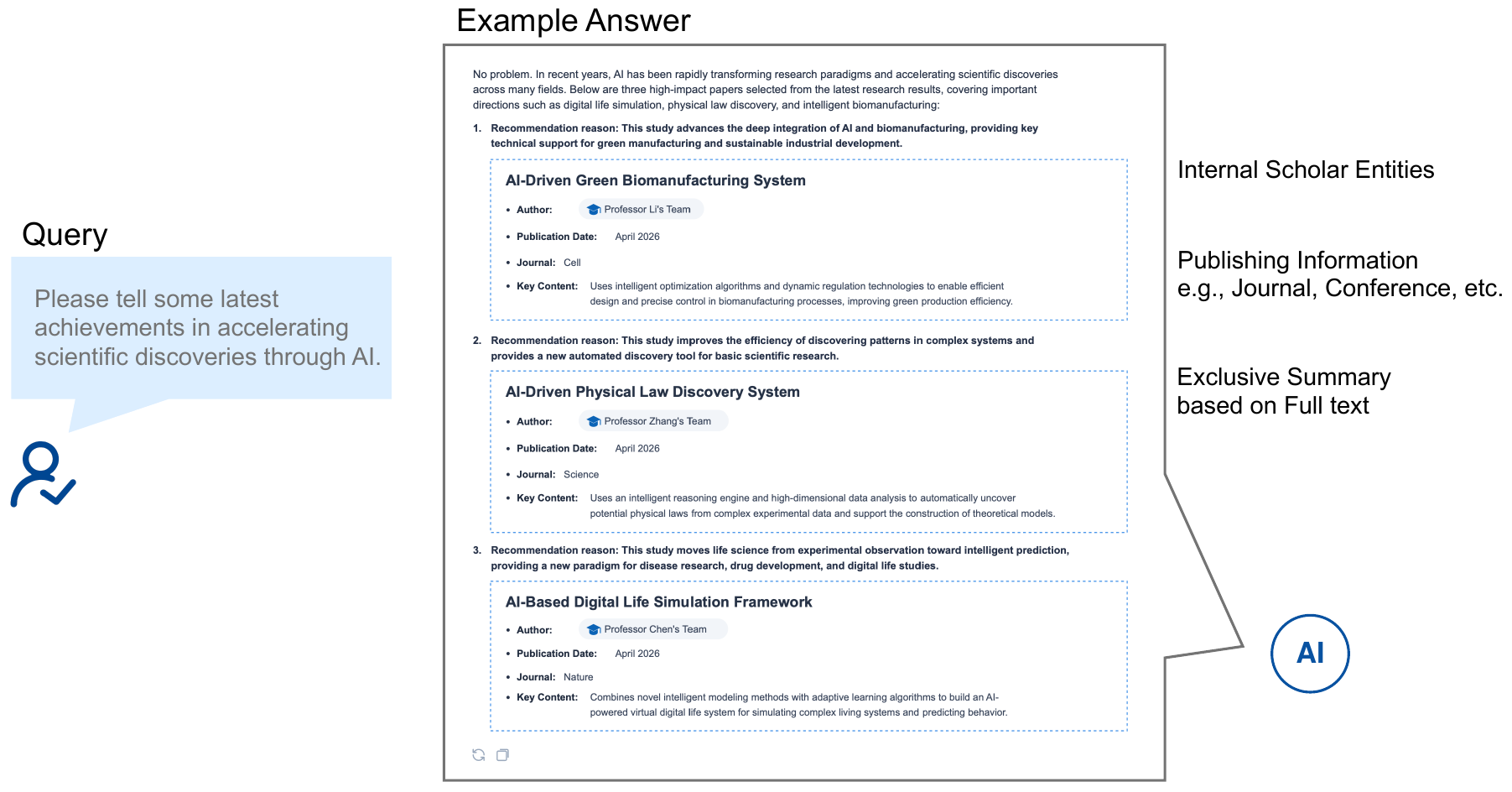}
  \caption{Research-direction-guided discovery returns related institutional papers with explanations and publication information.}
  \label{fig:direction-discovery}
  \Description{A research query is followed by relevant institutional papers, each accompanied by an explanation and publication information.}
\end{figure*}

As a collection grows, inspecting records individually becomes labor intensive, while either lexical or semantic matching alone can miss relevant work. Westlake Scholar's dialogue service exposes a paper-recommendation tool to the language model. When called, the tool runs BM25 and vector retrieval over approved institutional paper text, fuses the two ranked lists, and returns passages and identifiers from the top three papers as evidence. The model then explains the relevance of these candidates without searching an unrestricted external corpus (Figure~\ref{fig:direction-discovery}).

\subsection{Institutional Expert Discovery from Research Interests}

\begin{figure*}[t]
  \centering
  \includegraphics[width=\textwidth]{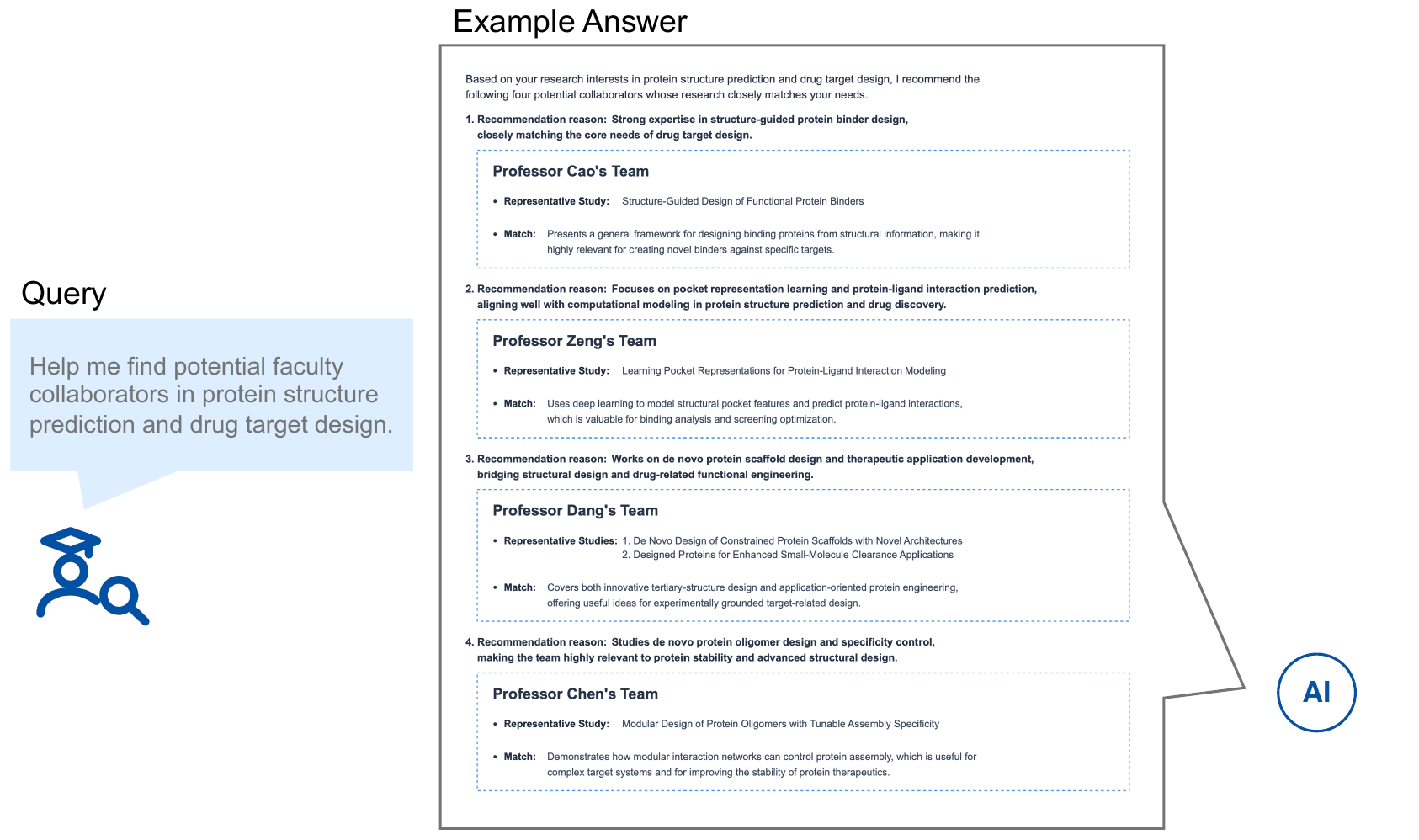}
  \caption{Institutional expert discovery connects a research-interest query to potential faculty collaborators and representative studies.}
  \label{fig:expert-discovery}
  \Description{A research-interest query leads to four potential faculty collaborators supported by representative studies and explanations of the match.}
\end{figure*}

Relevant expertise can be difficult to locate when a research interest crosses organizational boundaries and static directories expose only names and units. The expert-discovery service accepts a research direction, paper title, abstract, or research question. It incorporates recent user messages into the retrieval query, retrieves semantically related publication--scholar records, and instructs the language model to recommend only scholars supported by those records. Each suggestion cites representative publications as the basis for the match (Figure~\ref{fig:expert-discovery}). This publication evidence makes each suggestion inspectable and gives librarians and researchers a concrete basis for review and correction.

\subsection{AI-Generated Research Chronologies for Scholars}

\begin{figure*}[t]
  \centering
  \includegraphics[width=\textwidth]{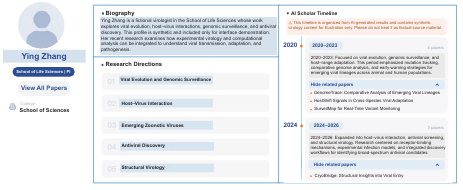}
  \caption{A research chronology for a scholar links period-level summaries to supporting publications.}
  \label{fig:scholar-timeline}
  \Description{A scholar profile presents research directions and a chronology with period-level summaries, topic tags, and expandable supporting papers.}
\end{figure*}

After a scholar has been identified, a publication list still gives limited support for understanding how that person's research has developed. Westlake Scholar groups linked publications into fixed five-year calendar windows. It checks DOI metadata against OpenAlex and Crossref, records unresolved years or conflicts for review, and prompts the configured language model to generate a 120--220-character focus summary and three to five tags from titles and abstracts. Longer periods are summarized in bounded evidence chunks and then synthesized. Each generated period retains paper counts, source-paper snapshots, model and prompt versions, and a fingerprint of its inputs. A chronology remains in a preview state until an authorized administrator publishes it, and readers can expand each period to inspect the underlying papers (Figure~\ref{fig:scholar-timeline}).

\section{Initial Institutional Deployment}

Westlake Scholar has operated at Westlake University since April 2026. The deployment established that the shared data foundation and four interfaces could operate together in a live institutional environment.

The deployment also places the platform within established information-science concerns about awareness, onboarding, and sustained faculty participation in IR services~\cite{Kim2011, Creaser2010}. These concerns informed a service process in which librarians and university users need accessible channels to identify needs, question generated content, and shape later iterations. This participatory-design commitment connects the technical integration to the continuing institutional work through which repository services are interpreted, governed, and improved~\cite{Davis2024, Lankes2008}.

\section{Ethics, Governance, and Limitations}

Expert recommendations and chronologies characterize identifiable faculty members, so provenance and correction matter beyond interaction-data privacy. Westlake Scholar admits papers to public search and recommendation only after institutional content review. Expert recommendations are generated on demand from retrieved publications and are framed as suggestions supported by named works. For chronologies, metadata conflicts and unresolved publication years are recorded, generated drafts remain non-public until administrative review, and every published period retains links to its source papers. Authorized administrators can reject or archive a generated chronology, and corrections to source records pass through the same institutional review workflow.

Governance also requires institutional procedures beyond the software. Until scholar-facing consent, correction, and opt-out controls are added, institutions publishing generated chronologies should provide notification and channels for correction, suppression, and appeal, and exclude generated summaries and expert rankings from personnel evaluation.

The ongoing institutional deployment provides a real-world setting for evaluating usage, retrieval quality, generated outputs, and scholars' responses, with the findings informing subsequent design and governance.

\section{Conclusion}

Westlake Scholar shows how an institution-controlled knowledge foundation can connect contextual reading, paper discovery, expert matching, and longitudinal views of scholarly work. By integrating repository content and scholar relationships with provenance, review, and institutional control, it extends the role of the institutional repository from deposit and access toward interactive scholarly discovery. The open-source implementation makes this architecture inspectable and adaptable, while the ongoing deployment provides a setting for continued evaluation and participatory governance.

\begin{anonsuppress}
\section*{Artifact Availability}

Westlake Scholar is the Westlake University deployment of Airalogy Scholar, an open-source platform developed by Hangzhou Airalogy Technology Co., Ltd., which retains copyright. The Airalogy Scholar source code and deployment documentation are available under the Apache License 2.0 at \url{https://github.com/airalogy/scholar}. Institutional records, deposited files, and usage or interaction data are internal university data and are not publicly released. A public demonstration is available at \url{https://scholar.airalogy.com}.
\end{anonsuppress}

\begin{acks}
We thank engineers Hongji Zhang and Zhao Wang and designer Qian Yu of Hangzhou Airalogy Technology Co., Ltd. for their contributions to the development and design of the system. We also thank Westlake University undergraduate students Haotian Jin, Qianlin Gu, and Zirui Zhao for their assistance with data collection.

\end{acks}

\bibliographystyle{ACM-Reference-Format}
\bibliography{sample-base}

\end{document}